\documentstyle[multicol,aps]{revtex}
\begin{document}
\draft
\title{SMALL ANGLE NEUTRON SCATTERING OF AEROGELS: SIMULATIONS AND
EXPERIMENTS}

\author{ Anwar Hasmy, Marie Foret, Eric Anglaret, Jacques Pelous, Ren\'e Vacher and
R\'emi Jullien}
\address{Laboratoire de Science des Mat\'eriaux Vitreux, UA 1119 CNRS,
      Universit\'e Montpellier II, Place Eug\`ene Bataillon,
               34095 Montpellier Cedex 5, France}

\date{\today}
\maketitle

\begin{abstract}
A numerical simulation of silica aerogels is performed using
diffusion-limited
cluster-cluster aggregation of spheres inside a cubic box (with periodic
boundary conditions). The volume fraction $c$ is taken to be
sufficiently large to get a gel structure at the end of
the process. In the case of monodisperse spheres, the wavevector
dependent scattered intensity $I(q)$ is calculated from the product of
the form factor $P(q)$ of a sphere by the structure factor $S(q)$,
which is related to the Fourier transform of $g(r)-1$, where $g(r)$ is
the pair correlation function between sphere centers. The structure factor
$S(q)$ exhibits large-$q$ damped oscillations characteristics of the short
range (intra-aggregate) correlations between spheres. These oscillations
influence the $I(q)$ curve in the $q$-region between the fractal
regime and the Porod regime and quantitative comparisons are made with
experiments on colloidal aerogels. Moreover, at small-$q$ values,
$S(q)$ goes through a maximum characteristic of large range
(inter-aggregate) correlations. Quantitative fits of the
maximum in the experimental $I(q)$ curves of base-catalyzed aerogels are
presented.  In the
case of polydisperse spheres, $I(q)$ is calculated directly from a single
aggregate simulation. It is shown that increasing polydispersity shifts
the location of the cross-over between the fractal and Porod
regimes towards
low $q$-value.
\end{abstract}
\begin{multicols}{2}

{\bf 1. Introduction}

As revealed by Small Angle X-rays Scattering (SAXS) or Small Angle Neutron
(SANS) experiments, Silica Aerogels 
\cite{smk,fks,vwp1} are made of a disordered, but homogeneous, array of connected 
fractal clusters which result from
the aggregation of primary particles \cite{jb}. In the case of colloidal
aerogel the form factor of the individual
particles is well defined and one can extract the 
precise form of the structure factor $S(q)$
from the scattered intensity $I(q)$. The analysis of the wave vector dependence of
$S(q)$, and $I(q)$, has permitted the determination of two characteristic
length scales which are the average size, $a$, of the  particles and the average size,
$\xi$, of the clusters.
Three distinct domains of wavevectors can be identified in both functions. 
At large $q$ ($q>a^{-1}$)
$S(q)$, and $I(q)$, exhibits large-$q$ damped oscillations, and the Porod law
($I(q)\sim q^{-4}$), respectively. At intermediate values of $q$, for
$\xi^{-1}<q<a^{-1}$, the fractal nature of intra-cluster particle
correlations is revealed by  a power law behavior $q^{-D}$, where $D$ is the
fractal dimension \cite{m} of the clusters. At last, at small $q$ values, for $q<<\xi^{-1}
$, the scattering function
saturate and eventually decreases as $q$ tends to zero.

Earlier studies \cite{dsw,f,gk,s,fpv} of the damped oscillations in the $S(q)$ curve of colloidal
aerogels have not been analyzed in terms of their all short range (intra-aggregate)
features. On the other hand,
the presence of a maximum in some $S(q)$ and $I(q)$ 
curves was puzzling for some authors \cite{dac,cdl} who tried
to fit the experimental data by considering single aggregate theories like the 
formula due to Fisher and Burford \cite{fb}
in which $\xi$ plays the role of a radius of gyration.
Others \cite{fks,vwp1,dsw,ffc}
have used a semi-empirical formula obtained by introducing a cut-off
function to limit the fractal scaling, in which $\xi$ enters as a correlation
length.  In all
cases, these approaches do not take into account the inter-cluster interactions
which give rise to a maximum in the scattering function. As far as
long range properties are concerned,
the single-aggregate approach is only valid for extremely diluted solutions
of aggregated particles where the mean inter-aggregate distance $\ell$ is
much larger than the
mean radius of gyration $R$ of the aggregates. Otherwise, the 
theoretical scattering function curve
of a single aggregate, which saturates for $q$ values 
smaller than $R^{-1}$, in the
so-called Guinier regime \cite{gf}, is not valid down to $q$
values of order $\ell^{-1}$,
where inter-aggregate correlations start to has some influence.
Recently, a hard sphere model have been introduced
by Posselt et al. \cite{ppm} to describe the packing of the connected clusters. 
However, their model does not take account of the fact that neglecting
other scattering contributions such as scattering by
thermally activated fluctuations,
should cause $S(q)$ and $I(q)$ to vanish for $q<\ell^{-1}$, since,
for distances larger than $\ell$, the
system becomes homogeneous and should no longer scatter the incident beam. Such behavior
cannot be avoided in the case of aerogels where $\ell$ and $R$ are  of the same order of
magnitude and should be replaced by $\xi$. If the experiments can be performed down
to sufficiently small $q$-values, if no other artifactual inhomogeneities are present and
if scattering by thermal fluctuations is small enough, all the experimental $I(q)$
curves for gels or aerogels should exhibit a maximum.
In practice, a maximum is observed,
or not, depending on  the range of $q$-values available.
When it exists, this maximum
is more or less pronounced, depending on preparation, i.e.
catalysis conditions.

This paper is a review of a series of recent numerical studies which
has been, in part, published elsewhere \cite{hfp,hvj,haf}. 
We first study the large-$q$ regime which is related to short
interparticle distances within the aggregate. Then, we present quantitative 
results for the location and shape of the maximum of the $I(q)$ curve at small-$q$
values. Finally, the effect of particle size polydispersity in the cross-over
between the fractal and Porod regime is discussed.
The scattering functions $S(q)$ and $I(q)$ were obtained 
from numerical simulations of the full aerogel structure which
is modelled by diffusion-limited cluster-cluster 
aggregation model (DLCA) in a box \cite{jb,me1,kbj}
with a sufficiently large initial concentration, rather than 
the hierarchical model \cite{bjk} which is only
able to build a single aggregate.

{\bf 2. Constraints on Theory}

{\it 2.1. The Model}

We have considered a three dimensional off-lattice extension 
of the original cluster-cluster aggregation model \cite{me1,kbj} in the case
of a sufficiently large particle concentration to get a 
gelling network at the end of the process. Such a
model was previously proposed by Kolb and Herrmann \cite{kh} to 
describe the formation of gels, but, at this time, they
considered a two-dimensional model on a lattice.
Initially, identical spherical particles of unit diameter are 
randomly disposed in a cubic box of edge length $L$ (here
$L$ is not necessary an integer) using a standard sequential 
addition procedure: attempts are made to center particles,
one after another, at points whose coordinates are random 
numbers uniformly distributed between 0 and $L$. If a particle overlaps
a previous one, it is discarded and a new trial is made. If the 
process generates $N$ particles, the dimension-less
concentration, or volume fraction, $c$ is given by:  

\begin{eqnarray}
c={\pi\over 6}{N\over L^3}
\label{E1}
\end{eqnarray}

From previous studies \cite{c}, it is known that, with this procedure, $c$
cannot exceed an upper limit which is called the ``jamming concentration'', 
$c_j = 0.385$. Let us consider the starting configuration  as a collection of
aggregates containing one particle each. At a later time, one obtains 
a collection of $N_a$ aggregates, the $i$-th aggregate containing
$n_i$ particles, so that:

\begin{eqnarray}
\sum_{i=1}^{N_a} n_i = N
\label{E2}
\end{eqnarray}

The algorithm proceeds as follows. An aggregate $i$ is chosen at 
random according to a probability, $p_{n_i}$ which depends
on the number of particles $n_i$ that it contains given by:

\begin{eqnarray}
p_{n_i} = {n_i^{\alpha}\over \sum_i
n_i^{\alpha}}
\label{E3}
\end{eqnarray}
In most of our simulations, we have taken $\alpha = - 0.55$, a
value close to $-{1\over D}$, where $D\simeq 1.78$ is the
fractal dimension of the resulting aggregates built in 3d \cite{me2},
in order to insure that the diffusion coefficient of
the aggregates varies with the inverse of their radius.
Then a space direction is chosen at random among the six directions 
$\pm 1, \pm 1, \pm 1$ and an attempt is made to move the cluster by
a step of one unit length in that direction (note that this choice
corresponds to perform a random translational brownian motion on a lattice 
but, since the original coordinates of the particles are not integers,
the aggregates themselves are built off lattice). If the cluster 
does not collide any other cluster during this motion, the displacement
is performed and the algorithm goes on by choosing again a cluster at 
random, etc... If instead a collision occurs, the cluster is translated 
in the chosen direction by the shortest distance insuring that one
of its particles becomes tangent to one particle of the collided cluster.
Then the collection of clusters is updated: the two colliding clusters
are discarded and a new cluster, formed by sticking together the colliding
clusters, is added to the collection. After that, one cluster
is chosen at random, etc... Periodic boundary conditions are used at the
edges of the box. The process is stopped when
a single aggregate is reached. If the concentration is larger than a
characteristic gel concentration $c_g$, the final aggregate spans
the box from edge to edge in the three space directions. This is the 
usual convention to define a gel network.

A series of calculations has been done to determine the gel
concentration $c_g$ as a function of the box size $L$. In practice 
we have varied the concentration and we have performed
twenty independent runs for each concentration. The gel 
concentration has been defined as being the concentration at
which ten runs end up with a gel. The results are given in 
figure 1 as a log-log plot of $c_g$ versus $L$. As already
found by Kolb and Herrmann \cite{kh}, the gel concentration tends 
to zero in the infinite $L$ asymptotic limit. Since, at the gelling
threshold, an aggregate of fractal dimension $D$ reaches the 
size $L$ of the box, one should have: 

\begin{eqnarray}
c_g \sim {L^{D}\over L^3} \sim L^{-(3-D)}
\label{E4}
\end{eqnarray}
Our data are well fitted with the slope $-1.28 \pm 0.05$, giving
$D \simeq 1.72\pm 0.05$, a value slightly smaller but quite close
to the fractal dimension $D=1.78$ of DLCA at the gel point in three
dimensions \cite{me2}.

In all the calculations reported below, we have considered $c$ values
much larger than $c_g$, in order to get a gel far above the gelling 
threshold and to be sure that the correlation length $\xi$ is smaller
than the box size $L$.  Typical examples of gel networks for two 
different concentrations are depicted in figure 2a and 2b where
we have visualized a two-dimensional projection of a slice of the
box. The slice thickness has been chosen to be proportional
to ${1\over c}$ in order to get the same mean coverage in 
projection. Thus the only difference between the two pictures is in
the way the apparent local densities deviate from the mean. As one sees
larger holes in case (a), it is apparent that the mean cluster size 
is larger in that case where the concentration is
smaller than in case (b). For comparison, we show in figure 2c a micrograph
of a colloidal aerogel sample.

{\it 2.2. The correlation function g(r)}

As usual the two points correlation function $g(\overrightarrow r)$ is 
defined such as $g(\overrightarrow r)d^3r$ is proportional to the
probability  of finding a particle center in a volume $d^3r$ at a 
distance $\overrightarrow r$ from a given particle center. Consequently,
for an isotropic material, the number of particle centers $dn$ located
between $r$ and $r+dr$ from a given particle center is proportional to 
$g(r)4\pi r^2dr$. Knowing from eq. 1 that, in average, the number of 
particle centers per unit volume is $6c\over\pi$, one can normalize $g(r)$
to unity when $r$ tends to infinity, by writing: 

\begin{eqnarray}
dn = {6c\over\pi}g(r)4\pi r^2 dr = 24 c g(r) r^2 dr
\label{E5}
\end{eqnarray}
We have used this formula to compute $g(r)$ in the gels resulting
from our simulations. For each particle in the box, we have counted 
the number of particle centers located between spheres of radius
$r$ and $r+\delta r$ taking care of the periodic boundary conditions
when investigating regions outside of the box. Then
we have averaged the result over
all the $N$ particles in the box and divided it by $24 c r^2 \delta r$.
Moreover, $g(r)$, obtained that way, has been
averaged over a large number of independent simulations.

{\it 2.3. The scattering functions S(q) and I(q)} 

From the single scattering theory, the structure factor $S(q)$
of a macroscopic system containing identical particles
with mean volume fraction $c$ is given by \cite{fs}:   

\begin{eqnarray}
S(q) = 1 +
{6c\over \pi}\int_0^{\infty} (g(r)-1){\sin qr\over qr} 4\pi r^2 dr
\label{E6}
\end{eqnarray}
The infinite boundary of the integral means that the incident beam
is scattered by the particles located within the whole macroscopic volume.
The presence of $g(r)-1$ means that one has subtracted the intensity
scattered by a quasi-infinite homogeneous object having the same
boundaries as the considered macroscopic system. As a consequence,
with formula (6), $S(q)\rightarrow 0$ when $q\rightarrow 0$ (because the
$q\simeq 0$ contribution of the boundaries has been suppressed). 
Quantitatively, this $q=0$ limit results from the following
sum rules: 

\begin{mathletters}
\begin{eqnarray}
{6c\over \pi}\int_0^{\infty}  g(r) 4\pi r^2 dr = M-1
\label{E7a}
\end{eqnarray}
{\rm and,}
\begin{eqnarray}
{6c\over \pi}\int_0^{\infty}  4\pi r^2 dr = {M}
\label{E7b}
\end{eqnarray}
\end{mathletters}
where $M$ is the total number of particles contained in the 
macroscopic volume. 

When applying formula (6) to real systems, it should
be remembered that the underlying theory considers only the single
scattering by static (quenched) particles. Therefore the  scattering 
by thermally activated (and correlated) motions of
these particles is not considered. It is known that in the case of 
liquids, the latter contribution (which there corresponds to scattering by
thermally activated density fluctuations) gives a non zero contribution 
at $q=0$ proportional to the product of temperature, bulk density
and isothermal compressibility \cite{e}. Since in the following,
the formula will be applied to solid aerogels, of quite low
compressibility, we will assume that such 
contribution is negligible.

In practice, we have numerically calculated $S(q)$ from the 
preceding $g(r)$ results, by replacing formula (6) by:

\begin{eqnarray}
S(q) = 1 +
{6c\over \pi}\int_0^{r_m} (g(r)-g_0){\sin qr\over qr} 4\pi r^2 dr
\label{E8}
\end{eqnarray}
Here, $g_0$ is a parameter which is very close, but not strictly equal,
to one, $r_m$ is an upper cut-off and the integral is numerically
computed as a discrete sum. We have chosen $r_m = {L\over 2}$, to 
avoid the boundary artifact mentioned above but also we have averaged 
$g(r)$ over many simulations and we have limited ourselves to
concentrations sufficiently greater than the gel concentration
to obtain a significant range of $r$ values  below $r_m$ where $g(r)\simeq 1$.
Nevertheless, the truncation of the integral implies
that the numerical results for $S(q)$ become meaningless for $q$
values smaller than:

\begin{eqnarray}
q_{\hbox{min}} = {\pi\over r_m}
\label{E9}
\end{eqnarray}
and, indeed, we have observed that for $q<q_{\hbox{min}}$ the precise 
shape of the computed $S(q)$ curve depends on both
$r_m$ and $g_0$. To avoid this problem, we have forced the sum 
rules (7a) and (7b) to be verified, and, instead of using
$g_0 = 1$, we have computed this parameter from: 

\begin{eqnarray}
g_0 = {\int_0^{r_m} g(r) 4\pi r^2 dr + {\pi\over 6c}\over\int_0^{r_m} 4\pi r^2 dr}
\label{E10}
\end{eqnarray}
where, to compute the integrals, we have used the same discretisation as
in formula (8). We have checked that $g_0$ is always equal to 1 within 
less that 0.001, however using (10) instead of
$g_0 = 1$, insures that $S(q)\rightarrow 0$, exactly,  when $q\rightarrow 0$.
This trick, which allows to obtain a continuation of $S(q)$ below
$q_{\hbox{min}}$, is expected to give a reasonably correct result if,
in the corresponding infinite system, $g(r)$ is supposed to stay constant 
for $r>r_m$. This has been checked {\it a posteriori} by verifying
that the numerical results are the same (within the numerical 
uncertainties) for different $L$-values.

For systems formed by monodisperse particles, the scattered intensity 
$I(q)$ can be obtained by using:

\begin{eqnarray}
I(q) = S(q)P(q)
\label{E11}
\end{eqnarray}
where $P(q)$ is the normalized form factor for spherical particles
of unit diameter:

\begin{eqnarray}
P(q) = (24{\sin{q\over 2} - {q \over 2}\cos{q\over 2}\over q^3})^2
\label{E12}
\end{eqnarray}

If the system contains polydisperse particles one  can no longer 
calculate the scattered  intensity $I(q)$  as  a  product of $P(q)$ and
$S(q)$.  One  should go back to the calculation  of  the
scattered amplitude \cite{fs},  which is proportional to:
\begin{eqnarray}
\tilde{A}=\sum\limits_i\int\limits_v{e^{i\vec{q}.(\vec{r}_i+
\vec{x})}{\rm d}^3x} \label{E13}
\end{eqnarray}
where $\vec{r}_i$ refers to the center of the $i$-th particle and $\vec{x}$  refers
to a running point inside the volume of the $i$-th particle with
respect  to  its  center. The integral inside the  sum,  which
should be performed over the volume of the $i$-th particle,  can
be  calculated  as  a  function of $a_i$, assuming  isotropy  and
homogeneity inside the sphere, leading to:
\begin{mathletters}
\begin{eqnarray}
\tilde{A}=\sum\limits_i{e^{i\vec{q}.\vec{r}_i}A_i(q)} \label{E14a}
\end{eqnarray}

{\parindent 0mm \rm with:}

\begin{eqnarray}
A_i(q)=4\pi{\sin({qa_i\over 2})-({qa_i\over 2})\cos({qa_i\over 2})\over {q^3}}\label{E14b}
\end{eqnarray}
\end{mathletters}
Then, assuming a random orientation of the aggregate over  the
direction  of $\vec{q}$,  the  scattered intensity  $I(q)=|\tilde{A}|^2$ can  be
written as:
\begin{eqnarray}
I(q)=\sum\limits_{i,j}{A_iA_j{\sin(qr_{ij})\over qr_{ij}}}
\label{E15}
\end{eqnarray}
where $r_{ij}=|\vec{r}_i-\vec{r}_j|$.

 Note  that, since the $i$ and $j$ dependent product $A_i A_j$ appears
inside  the sum, the result cannot be split in two  parts.  In
particular  one cannot use the correlation function $g(r)$
to  calculate an intermediate structure factor $S(q)$. Here  the
double sum should be calculated directly. We have used such formula
to calculate $I(q)$ for single aggregates built using a hierarchical model.

{\bf 3. Results}

In figure 3 we show  typical $g(r)$ curve that 
result from an average over 20 simulations  with $L=57.7$ and $c=0.05$.
In this figure one observes a strong peak at $r=1$, a discontinuity at
$r=2$, and a weaker singularity at $r=3$ (discontinuity of the derivative).
At large $r$, g(r) goes through a minimum and becomes very close
to one (in order to avoid artifactual correlations due to the
periodic boundary conditions, we have used
$c$ values bigger than $c_g$ to insure that the the homgeneous regime
(when $g(r)$=1) can be reached for $r < {L \over 2}$).
For small $c$ values, one can observe the fractal regime, at intermediate $r$
values, where $g(r)$ follows the power law:
\begin{eqnarray}
           g(r) \propto r^{-(3-D)}
\label{E16}
\end{eqnarray}
The range of this fractal regime decreases as $c$ increases. 
Note that in the inset in figure 3, where $c=0.1$, it is almost inexistent.

The strong peak at $r=1$ is due to the non-zero proportion of distances $r=1$
corresponding to bonds between contacting particles. 
The non-zero value of $g(r)$ for $r=1_+$ and the discontinuity at
$r=2$ can be understood if one considers that $g(r)$ can be written as:

\begin{eqnarray}
       g(r)=g_1(r)+g_2(r)         
\label{E17}
\end{eqnarray}
where $g_1(r)$ is the contribution of couples of particles that are tangent to
the same third one and where $g_2(r)$ contains all the other
contributions. We have observed that, while $g_2(r)$ is continuously varying
from $r=1$, where $g_2(1) = 0$, up to the largest distance,
going through a maximum around $r=2$, $g_1(r)$ exists only between $r=1$ and
$r=2$ and reaches non-zero values at both limits. The weaker singularity at $r=3$
can be explained by analyzing distances between spheres
tangent to each sphere of a dimer 
in a manner very similar to the discontinuity at $r=2$.

In the inset in figure 3 we show the $g(r)$ curves obtained for $L= 57.7$ and 
different $c$ values by emphasizing the region near the minimum.
One can see that the location of the minimum strongly depends on $c$.
The minimum corresponds to distances between particles
located at the periphery of the clusters where the local 
density is smaller. Therefore the value of $r$ at the minimum gives
a good estimate of the mean cluster size, let us call it $\xi$.

The dependence of $\xi$ with $c$ is reported in the log-log
plot of figure 4. Assuming that the clusters are fractal with fractal
dimension $D$, one should get:

\begin{eqnarray}
\xi \sim c^{-{1\over 3-D}}
\label{E18}
\end{eqnarray}
The value  $\xi=3$ reported for the largest concentration $c=0.1$ should
be considered as artifactual since for such concentration the
minimum of $g(r)$ sticks on the singularity at $r=3$, as seen in the inset
of figure 3. A straight line fit of the remaining points for the largest 
box size gives a slope of $-0.77\pm 0.03$, leading
to $D = 1.70\pm0.06$, again slightly smaller but quite close to
the fractal dimension $D=1.78$ of 3d DLCA aggregates.
We point out that in this model $D$ increases when $c$ increases
as showed elsewhere \cite{vpd,hj}. However, the above value $D$ can be 
understood by analyzing the time evolution during the aggregation process
at the instant where the system can be
considered as a gelling network the fractal dimension of the clusters
is $D \approx 1.78$ and the characteristic length $\xi$ value does not change
compared with to measured $\xi$ value at the end of the aggregation
process \cite{hj}.

Typical $S(q)$ curves are reported in
figure 5 for different concentrations and for $L=57.7$. The location of
$q_{\hbox{min}}$ is indicated by the arrow. All the curves exhibit the 
same damped oscillations at large $q$. In this figure, one observes a 
large minimum at about $q\simeq 4$ followed by
damped oscillations. The oscillations can be attributed to the
$\delta$-peak of $g(r)$. If one considers only the distance contributions
of the delta peak the correlation function $g(r)$ writes:

\begin{mathletters}
\begin{eqnarray}
g_1(r)={z \over 24c} \delta (r-1) 
\label{E19a}
\end{eqnarray}

{\parindent 0mm \rm and its Fourier transform:}

\begin{eqnarray}
S_\infty(q)=1+z{\sin{q} \over q} 
\label{E19b}
\end{eqnarray}
\end{mathletters}
where $z$ is the coordination number of the system. In DLCA aggregates
$z=2$ and the corresponding $S_\infty(q)$ curve is depicted by 
the dotted line  in figure 5.
As expected, this approximation corresponds to the asymptotic large-$q$
limit of $S(q)$. However the large minimum at $q\simeq 4$ is not accounted for by this
contribution only. Not only the delta-peak at $r=1$ 
influences the shape of the first minimum of $S(q)$, but also
the others short range features of the system, as 
the discontinuity at $r=2$ and the singularity at $r=3$, play some role. 

The linear fractal regime is quite narrow
and is more extended for low concentrations. It corresponds to a fractal 
dimension $D\simeq 1.7$. We point out that in a system of homogeneous 
connected fractal clusters of finite size, the fractal dimension
is only an ``apparent fractal dimension'' as showed elsewhere \cite{hj,afl}.
At lower $q$ values, all the curves exhibit a maximum. The location of the
maximum, $q_m$, as well as the intensity of the maximum, $S(q_m)$, have 
been reported as a function of $c$ in table I together
with the corresponding values of $\xi$. If one forget 
the artifactual situation $c=0.1$, the product $q_m\xi$ is
almost constant, so that one has approximately: 

\begin{eqnarray}
q_m\simeq {2.75\over\xi}
\label{E20}
\end{eqnarray}
On the other hand one observes
that the intensity of the maximum $S(q_m)$ is roughly proportional 
to $c\xi^3$, which is the number of particles contained
in a sphere of diameter $\xi$:

\begin{eqnarray}
S(q_m)\simeq 0.75 c\xi^3
\label{E21}
\end{eqnarray}

For very low $q$-values (smaller than $q_m$) we get a linear behavior 
with slope $+2$. This is consistent with the low-$q$
expansion of formula (7), which predicts: 

\begin{mathletters}
\begin{eqnarray}
S(q)\simeq \zeta^2q^2\
\label{E22a}
\end{eqnarray}
with:
\begin{eqnarray}
\zeta^2 = {6c\over \pi}\int_0^{r_m}(g_0- g(r)) 4\pi r^4 dr
\label{E22b}
\end{eqnarray}
\end{mathletters}
This result is derived from a Taylor expansion of $\sin qr / qr$
inside the integral of (8). Note that, for an infinite system 
(where $r_m=\infty$ and $g_0=1$), such
procedure is mathematically justified if $g(r)-1$ tends to zero more 
quickly than any power law when $r$ tends to infinity.
Our numerical results, which leads to a quite size-independent 
result for $\zeta$, strongly support a large-$r$
exponential decay for $g(r)-1$. This defines
another characteristic length $\zeta$ which can be usefully compared to $\xi$.
In table I, we have reported the numerical estimates
of $\zeta$ from our computed $S(q)$ curves for the different $c$ 
values considered. We find that $\zeta$ is roughly proportional to $\xi$,
and therefore $\zeta$ does not bring any new information on the
structure.

As it should be expected, the characteristics of the maximum of $S(q)$
do not depend only on the mean cluster size $\xi$, but also on the extension 
of the cluster size distribution. The cluster size distribution
is reminiscent of the  size distribution of the aggregates observed 
during the aggregation process
and therefore it strongly depends on the nature of the aggregation mechanism.
As a consequence, all the quantitative analysis done above,
as well as the values of the constants appearing in formulae (20) 
and (21), are only valid for DLCA
where it is known that the aggregate size distribution presents a well 
defined maximum.     

When the cluster perform a ballistic random motion instead of 
diffusion brownian during the aggregation process described in section
2.1., one obtains the 
ballistically-limited cluster-cluster aggregation \cite{bj} (BLCA) model
which is able to describe the aerosols in the Knudsen regime.
In BLCA the polydispersity
of cluster sizes is larger than in the DLCA model. On the other
hand, when a very small sticking probability is introduced in the DLCA model
one obtains the chemically-limited cluster-cluster 
aggregation \cite{jk,bb} (CLCA) model where it is known that the aggregate
size distribution is broader than the DLCA and BLCA models. In figure
6 we show that this polydispersity features are observed in the $S(q)$
curves, where the broadest peak at $q_m$ corresponds to the CLCA model.

The log-log plots of $I(q)$ versus $q$ for different concentrations are
reported in figure 7. As a result of the multiplication by $P(q)$
and the use of the logarithmic scale, the maximum appears
to be relatively less pronounced than in the $S(q)$ curves. The large-$q$
oscillations in the Porod region are due to the monodisperse particles 
forming the system. These oscillations become more and more damped when
one increases 
the particles polydispersity \cite{hvj} as it will be showed in figures 10.

{\bf 4. Discussion}

In this section we would like to discuss the theoretical results in the 
light of experimental data.

{\it 4.1. Short-range correlations (large q-values)}

Colloidal silica aerogels have been prepared using the process described
in ref. \cite{fpv}.
They have densities ranging from 0.070 to 0.380 $g/cm^3$. They are made of
small colloidal
spherical particles with a quite low diameter polydispersity as it has been
checked on electron micrographs such as figure 2c. We have checked that
the intensity scattered by the correpondly diluted sol can be fitted by an 
averaged form factor $\overline{P(Q)}$ given by:

\begin{mathletters}
\begin{eqnarray}
\overline{P(Q)} = \int_0^\infty P(Q,a) g(a) da
\label{E26a}
\end{eqnarray}
{\rm where $P(Q,a)$ is the form factor of a spherical particle of diameter $a$
and $g(a)$ a truncated gaussian distribution given by:}
\begin{eqnarray}
g(a) \sim e^{-{1\over 2}({a-a_0\over\sigma})^2}
\label{E26b}
\end{eqnarray}
\end{mathletters}
The same analysis has been done
for all our
experimental data and in the following we give the results for the structure
factor as being the ratio of $I(Q)$ by $\overline{P(Q)}$ as a function of
the reduced wavevector $q=Qa_0$. All the experimental $S(q)$ curves,
determined this way, have been normalized such that $S(q)\rightarrow 1$ for
$q\rightarrow\infty$. It has been shown \cite{hfp} that using formula (11)
in presence of small polydispersity, in spite of this approach, the $S(q)$ 
curves present the same broad 
minimum.

In figure 8a one gives $\log S(q)$ versus $\log q$ for different aerogels made
of particles of the same size ($a_0 =96 \AA\ $) but with
densities ranging from 0.070 to 0.250 $g/cm^3$. On
this figure, one observes the same characteristic broad minimum followed by
damped
oscillations that we have observed in the simulated curves, and the density fixes the size $\xi$ of the
clusters.    
In figure 8b we compare two experimental $S(q)$  curves for the same aerogel
density ($\rho=0.10 g/cm^3$) with the simulated curve in the diffusion limited
case. It is clear in this figure that the agreement between
theory and experiments is only qualitative.
Even if the data are very noisy for
large $q$ values it seems that the large $q$ oscillations of the experimental
curves are
more damped. But we would like to focus on the larger discrepancy which is
that the minimum is wider and deeper in the experimental
curves. This discrepancy is systematically
more important for bigger particles, and its cannot be attributed to the
kind of restructuring as showed in ref. \cite{hfp} 
because a restructuring mechanism
in DLCA aggregates  increases its coordinations number resulting in the
$S(q)$ curves a deeper minimum but not wider.
Since it is difficult to imagine some other realistic restructuring     
mechanisms able to fully account for the observed discrepancies, we do not
trust the earlier interpretations
which were considering quite large coordination numbers \cite{dsw,f,gk,fpv}.
One might invoke other possible explanations for
the discrepancies such as small-$q$ modifications of the form
factor or corrections
to the scattered intensity due to some shape deformation of the
particles near
their contact zone. This last effect might be approximately taken into account
by considering a different length for the particle-diameter and for the
center-to-center
distance between contacting particles. However all these considerations, if
they might sometimes give a better fit, appear to be too ``ad hoc''
to really improve the comprehension of the problem.

Here we would like to propose another
tentative interpretation.
In general a complete theory of scattering (including multiple
scattering, shadowing, refraction etc...) should consider two dimensionless
parameters, $Qa$ and
$ka=2\pi{a\over\lambda}$.  The fact that the theoretical
$S(q)$ curve considered above does not depend
on the extra parameter $ka$ comes from all the considered approximations.
However the simple scattering theory should be recovered in
the limit $ka\rightarrow 0$. Some corrections might appear for large $ka$
values.
The fact that in figure 8b the theoretical curve can be considered as
the limit of the experimental ones when $a\rightarrow 0$ support this
analysis. Moreover
the parameter $ka$ is quite large in our case. We used a combination of
two incident  neutron
wavelengths of $6\AA\ $ and $18\AA\ $ in the experimental set-up. Thus our $ka$
values are in the range 30 to 300,
close to the values involved in the geometrical optics approximation.
It is
reasonable to admit that corrections to the simple scattering theory, such as
shadowing, refraction, multiple scattering effects, cannot be neglected for
such
large values.

{\it 4.2. Long-range correlations (small q-values)}

As is known, colloidal aerogels are formed by particles bigger than
standard aerogels. This fact impedes
to obtain information about its long-range correlations (small $q$- values).
For study the long range correlations we have considered SANS 
experiments on standard silica
aerogels \cite{vwp1,vwp2}. They are prepared 
by chemical reactions (hydrolysis and condensation)
of organosilicates. According to the $p_H$ value of the
hydrolysis aqueous solution, we can distinguish ``basic''
and ``neutral'' aerogels. The basic aerogels \cite{vwp2} are made of
larger sized, but strongly polydisperse, primary particles
while, for the neutral aerogels \cite{vwp1}, particle sizes are smaller
and extend down to the atomic scale. According to previous studies \cite{vwp1,vwp2,fpv},
only colloidal and basic aerogels can be considered as grown 
according to DLCA while the neutral aerogels are more likely
grown according to the chemically-limited cluster-cluster aggregation process
\cite{jb,jk,bb}. Three experimental $I(q)$ curves for
basic aerogels with different densities are compared with simulations
in figure 9. The
concentrations that we have used for the fit are those corresponding to
the aerogel density $\rho$, according to the formula:

\begin{eqnarray}
c = {\rho\over\rho_0}
\label{E23}
\end{eqnarray}
where $\rho_0 =2,2$g.cm$^{-3}$ is the density of silica. The only
two adjustable parameters are the mean particle diameter
value and a multiplicative constant for the intensity. Note
that varying  these parameters in a log-log plot does
not change the shape of the curve but only leads to translations
along the $x$ and $y$ directions. The three fits have
been performed together using the same values for these parameters.
A discrepancy occurs in the Porod region where the experimental
curves stay slightly below the maxima of the large-$q$ oscillations
of the theoretical curves. Such discrepancy
can be attributed to the strong polydispersity of the primary
particles \cite{hvj} and will be discussed in section 4.3..

We point out that, 
our DLCA model neglects rotational diffusion, aggregates
deformations as well as all kinds of restructuring effects.
However, as mentioned above, large
restructuring is certainly not present, at least in the early stages
of the aggregation process. But, in the last stages, when the gel
structure is under formation, the diffusion process looses its
leading character and it might be that restructuring and rotational
motions have some influence on
the gel structure and the inter-clusters correlations.

{\it 4.3. Effect of the particle size polydispersity}

To study the influence of the particle size polydispersity we have calculated
$I(q)$ , using formula (15), for hierarchical aggregates made of spheres whose 
diameter are distributed according to a truncated gaussian 
distribution characterized by $\sigma_{eff} = {\sigma \over a_0}$, where $a_0$
and $\sigma$ are the mean value and the standard deviation of their diameters \cite{hvj}.

It has been shown \cite{hvj} that the crossover wave-vector $q_c$
between the fractal and Porod regime, for $\sigma_{eff}>0.1$, varies as:

\begin{eqnarray}
q_c={2\pi \over a_0}(1-1.6 \sigma_{eff})
\label{E30}
\end{eqnarray}
where $\sigma_{eff}$ is the dimensionless standard deviation, 
or polydispersity. Note that, when $\sigma_{eff}=0$ the crossover wave-vector
$q_c$ is equal to ${2\pi \over a_0}$, as can be verified in figure 7. In figure 9 the experimental
curves exhibits the crossover take place at $q_c$ approximately equal
to $4\over a_0$ and according the 
above equation $\sigma_{eff}$ should be equal to $0.25$.
We have computed, for simplicity, a single
aggregate with a particle number $N$ equal to 32 using the hierarchical procedure \cite{bjk}
with a particle polydispersity $\sigma_{eff}=0.25$, the theoretical
$I(q)$ curve has been calculated using equation (15) and has been averaged over 32 simulations. The resulting
curve is shown in figure 10a (solid line) in a log-log plot of
$I(q)q^4$, emphasizing the large-$q$ region, and for comparison we have
depicted the case when $\sigma_{eff}=0$(dashed line) and the experimental
curve for $c=0.043$(open circles). We can note in this figure that
introducing a polydispersity not only the large-$q$
damped oscillations disappear, but also $q_c$ shifts towards
low-$q$ values. This result can be attributed to the fact that
larger particles dominate the scattering. Even if a single aggregate does not 
account for the realistic long-range correlations, 
figure 10a shows the good agreement at large $q$-values between the  
theoretical $I(q)$ curve
for $\sigma_{eff}=0.25$ and the experimental ones. Figure 10b
shows a log-log plot of $I(q)$ versus $q$. Here 
the dashed line represents the numerical $I(q)$ results
at the low-$(q)$ values in a system of connected 
fractal clusters for $c=0.043$,
and solid line represents results obtained at large-$q$ values of
a single aggregate made of polydisperse particles. In this figure 
the agreement between the simulation and the entire experimental 
results (open circles) is remarkable. 

{\bf Conclusion}

In this paper we have shown that the diffusion-limited cluster-cluster
aggregation model (DLCA) can explain the structure of ``colloidal''
as well as ``basic'' aerogels, as revealed by small-angle
neutron scattering experiments since in most cases the full $I(q)$
curve can be quantitatively accounted for. This is a real progress 
compared to previous approaches which were focusing on the 
intermediate-$q$ fractal regime only. Such modelization
is now used to study some other physical properties
of aerogels. In the present issue we give two examples
of application which are numerical calculations of aerogel
sintering \cite{joh} and simulations of phase transitions
in the pores of aerogels \cite{uhj}. We also present a numerical
study of the evolution of the $I(q)$ curve during the aggregation
process \cite{hj}. Other applications are under progress. Further
investigations are needed to account for the experiments on ``neutral''
aerogels which exhibit a larger fractal dimension  than DLCA aggregates.
Neutral aerogels are made of very small and strongly polydisperse
particles whose sizes extend down to the atomic scale. It might
be that their growing mechanism is closer to chemically-limited
aggregation (CLCA) than DLCA but also there might exist some complex
restructuring mechanisms due to their flexibility.

One of
us (A. H.) would like to acknowledge support from CONICIT (Venezuela).

\end{multicols}{2}

\begin{table}
\begin{tabular}{cccccccr}
$c$   & $\xi$ & $q_m$  & $q_m\xi$& $S(q_m)$ & $S(q_m)\over c\xi^3$ & $\zeta$ & ${\zeta\over\xi}$ \\
\tableline
0.025           & 10.7 & 0.26   &    2.78  & 24.7 &        0.81& 38.5 & 3.6       \cr
0.033           &  8.4 & 0.31   &    2.60  & 16.0 &        0.82 & 28.1 & 3.3      \cr
0.038           &  7.8 & 0.37   &    2.88  & 12.8 &        0.71 & 25.9 & 3.3      \cr
0.043           &  6.7 & 0.40   &    2.68  & 10.5 &        0.81 & 21.0 & 3.1      \cr
0.050           &  6.3 & 0.43   &    2.71  &  8.3 &        0.66 & 18.4 & 2.9      \cr
0.061           &  5.3 & 0.52   &    2.76  &  6.2 &        0.68 & 15.5 & 2.9      \cr
0.100           &  3.0 & 0.80   &    2.40  &  3.3 &        1.22 & 10.0 & 3.3      \cr 
\end{tabular}

\bigskip
\caption{For each concentration $c$ considered in the simulations, we have reported the location
$\xi$ of the minimum of $g(r)$, the location $q_m$ of
the maximum of $S(q)$,  the product
$q_m\xi$, the intensity of the maximum $S(q_m)$ and the ratio ${S(q_m)\over c\xi^3}$, the characteristic
length $\zeta$ entering the low-$q$ expansion of $S(q)$ and the ratio
${\zeta\over\xi}$. We recall that $q$ is here a dimensionless quantity which
is, in fact, equal to
$2Qr_0$ where $Q$ is the dimensioned wavevector and $r_0$ is the radius of
the primary particles
and also that $S(q)$ has been normalized to unity for large $q$.}
\end{table}


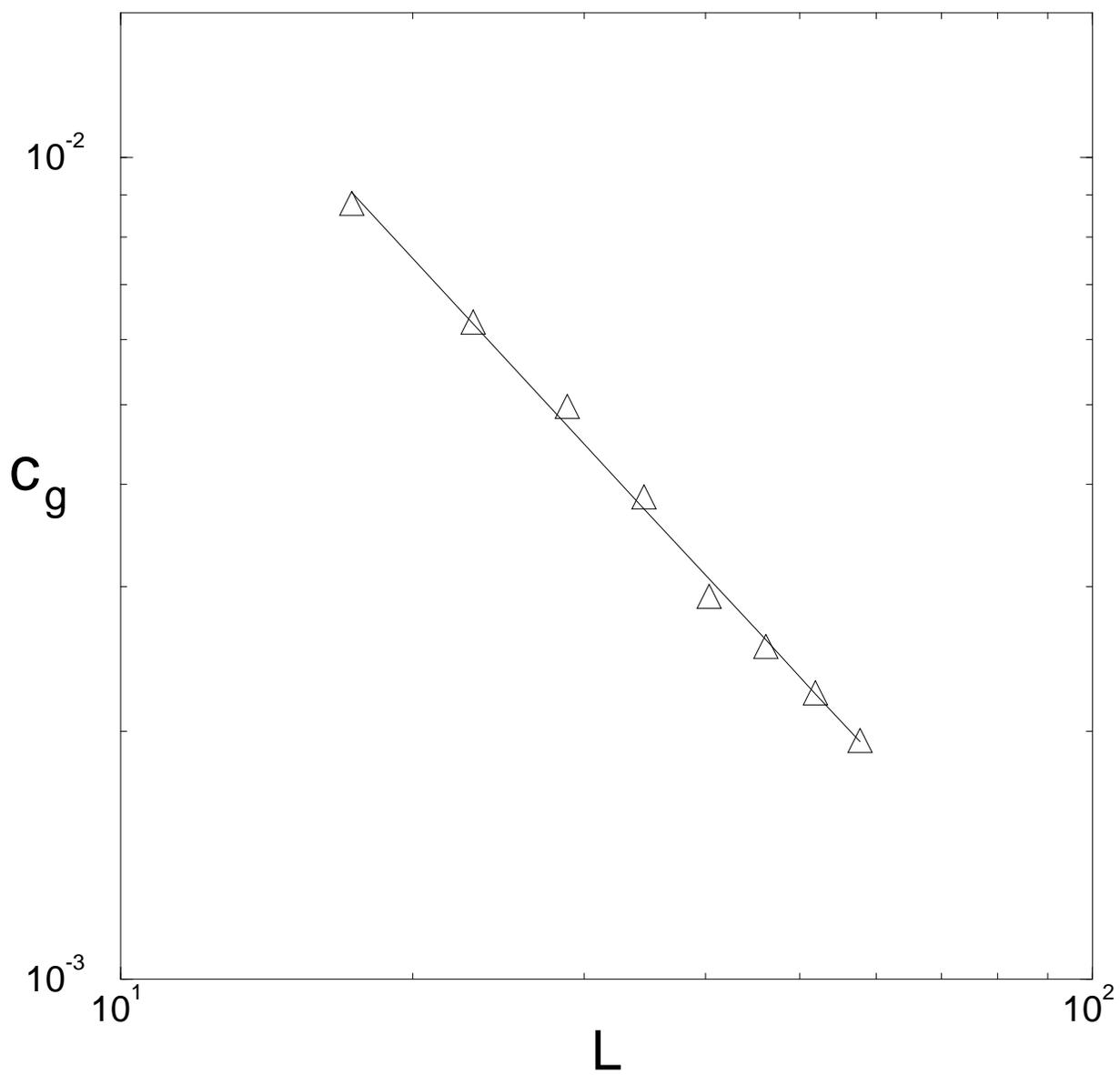
\begin{figure}
\caption{Log-log plot of the gel concentration $c_g$ as a function of the
 box size $L$. The fit by a straight line is shown which gives a slope of $-1.28
\pm0.05$.
}
\end{figure}

\begin{figure}
\caption{Two dimensional projections of the particles contained in  a
slice of thickness $\ell$ after obtaining a gel in a box
of size $L=57.7$. Cases (a) and (b) corresponds to $c=0.0095, \ell=34.6$ and 
$c=0.038, \ell=8.65$, respectively. Cross sections of particles that
are cut by the
front slice edge are shown in black. (c) Micrograph of a 270 $\AA$ colloidal
aerogel sample.}
\end{figure}

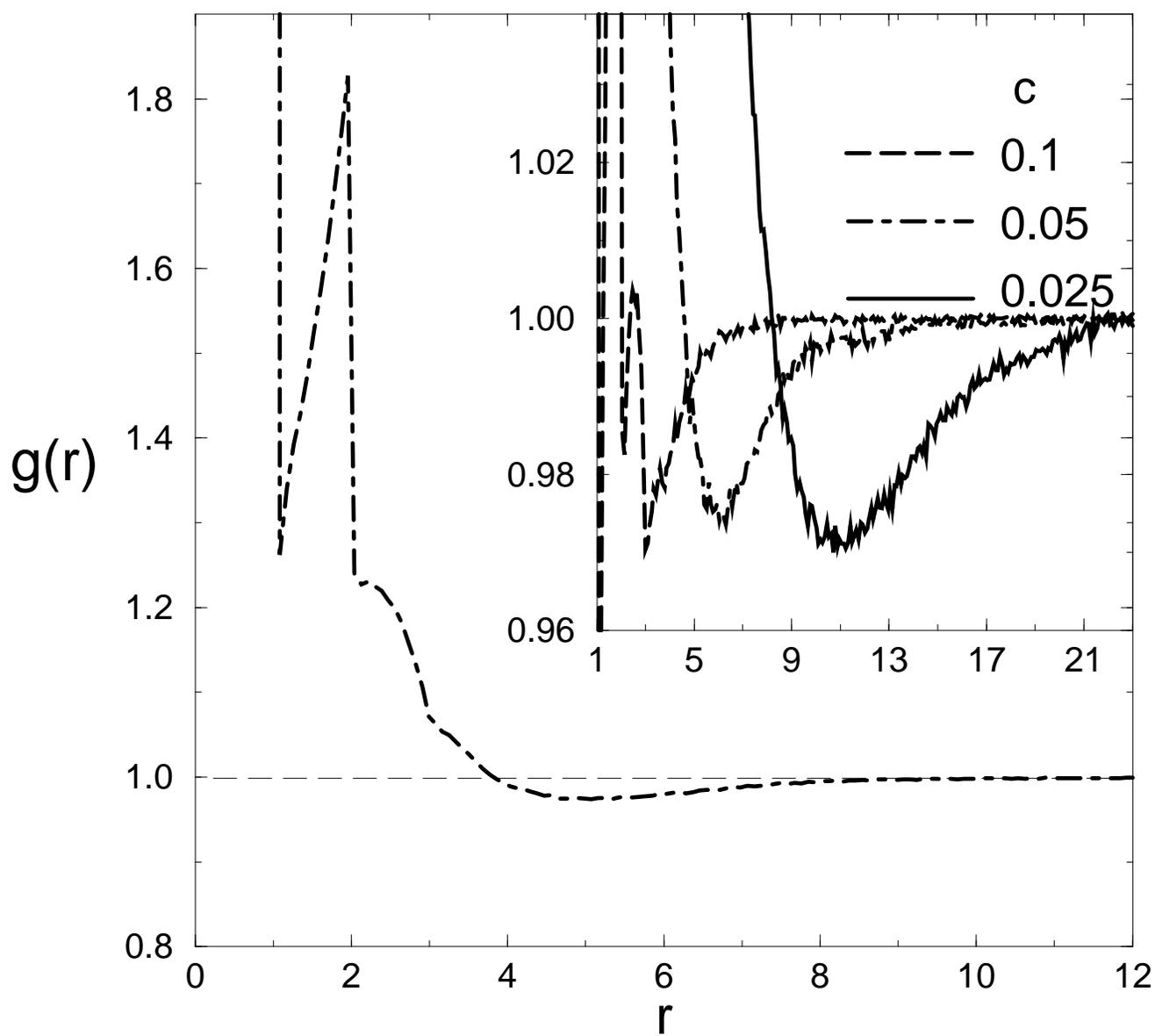
\begin{figure}
\caption{ Plot of $g(r)$ versus $r$ for $L=57.7$ and
$c=0.05$. Inset: g(r) curves for different $c$ values ($c=0.025, 0.05, 0.1$).
These curves result from averages over 20
simulations.}
\end{figure}

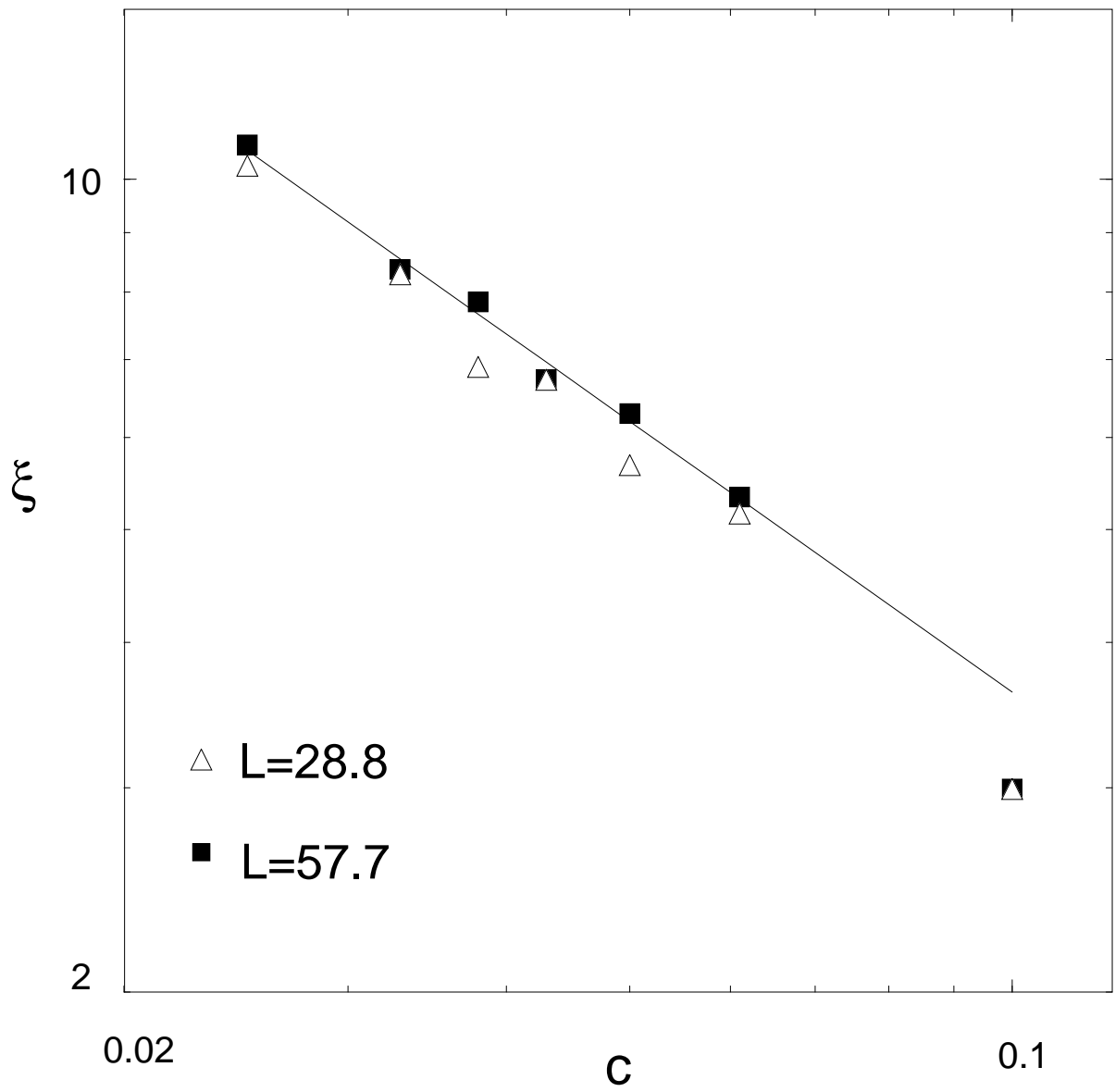
\begin{figure}
\caption{Log-log plot of $\xi$, location of the
minimum of $g(r)$, versus $c$, for $L=28.8$ and $L=57.7$. The fit by a
straight line of the $L=57.7$ data, excluding the $c=0.1$ point, is shown giving
a slope of $-0.77\pm0.03$.}
\end{figure}

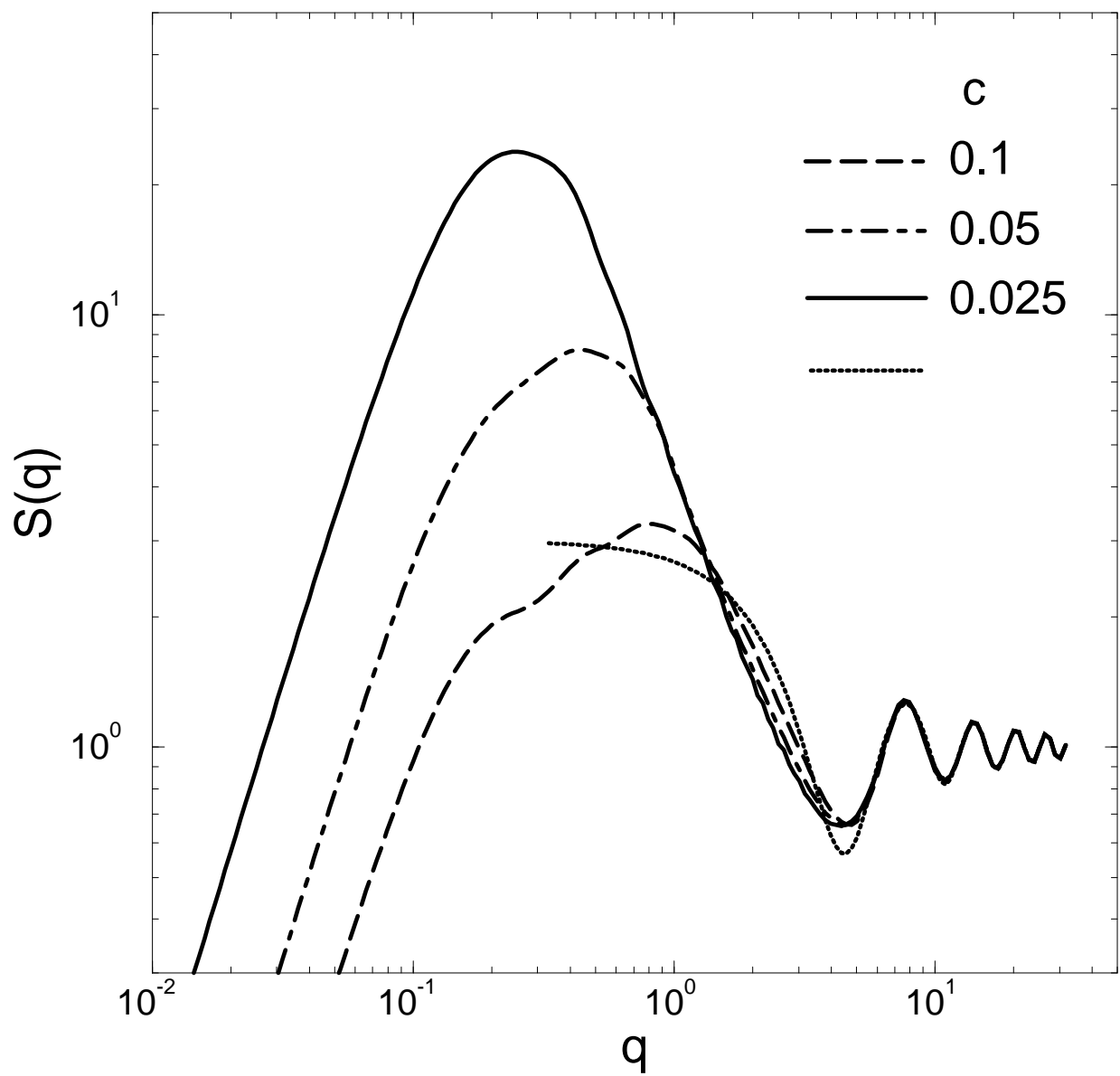
\begin{figure}
\caption{$S(q)$ versus $q$  for $L=57.7$ and different $c$ values:
c=0.025, c=0.05, c=0.1. The dotted line correspond to $S_{\infty}(q)$. Here and in the following figures the location of 
$q_{\hbox{min}}$ is indicated by the arrow.}
\end{figure}

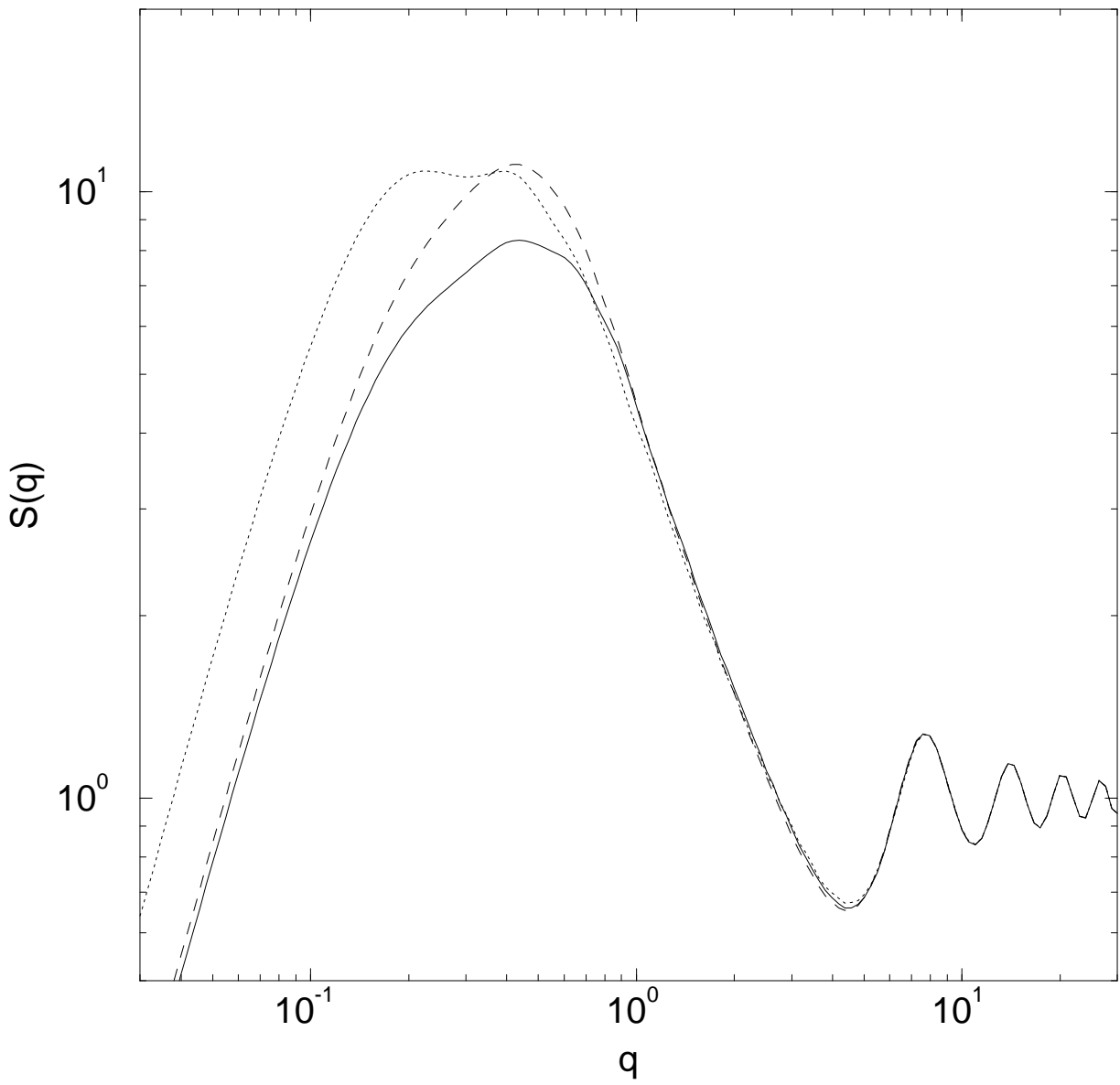
\begin{figure}
\caption{Log-log plot of $S(q)$ versus $q$ for $L=57.7$ and $c=0.05$,  the solid line,
dashed line and dotted line, correspond to DLCA, BLCA and
CLCA model, respectively. These curves result from average over 20 simulations.}
\end{figure}

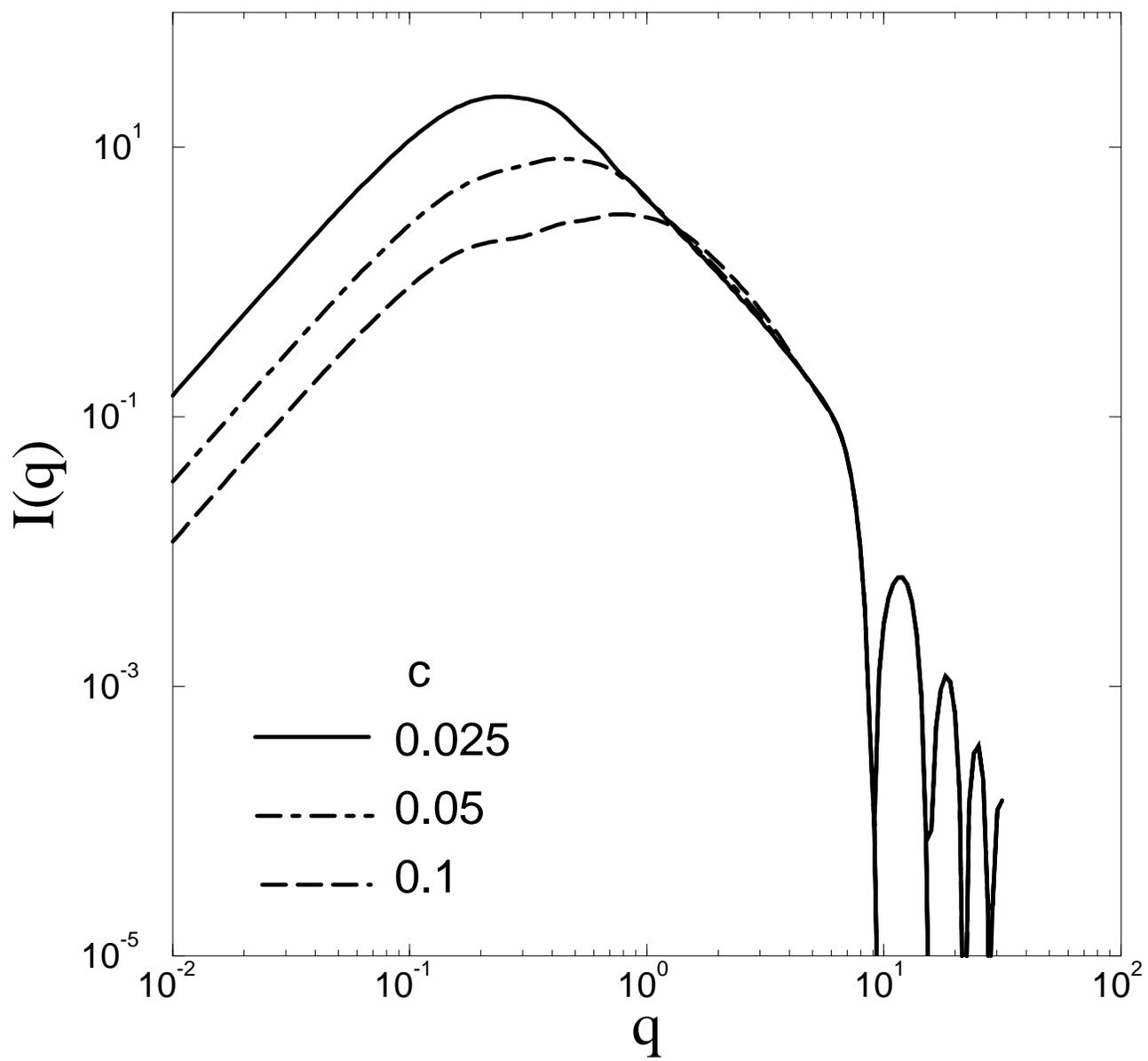
\begin{figure}
\caption{Log-log plot of the scattering intensity curve $I(q)$ versus $q$
for different $c$-values. The parameters are the same as in figure 5.}
\end{figure}

\begin{figure}
\caption{
(a) Experimental $S(q)$ curves (obtained as explained in text) for
the 96$\AA\ $-colloidal aerogel family. Samples are labeled by their densities. 
(b) Experimental $S(q)$ curves for  samples of various particle diameters 
but with the same density, $\rho = 0.10 g/cm^3$,
and the simulated $S(q)$ curve in the diffusion-limited case.}
\end{figure}

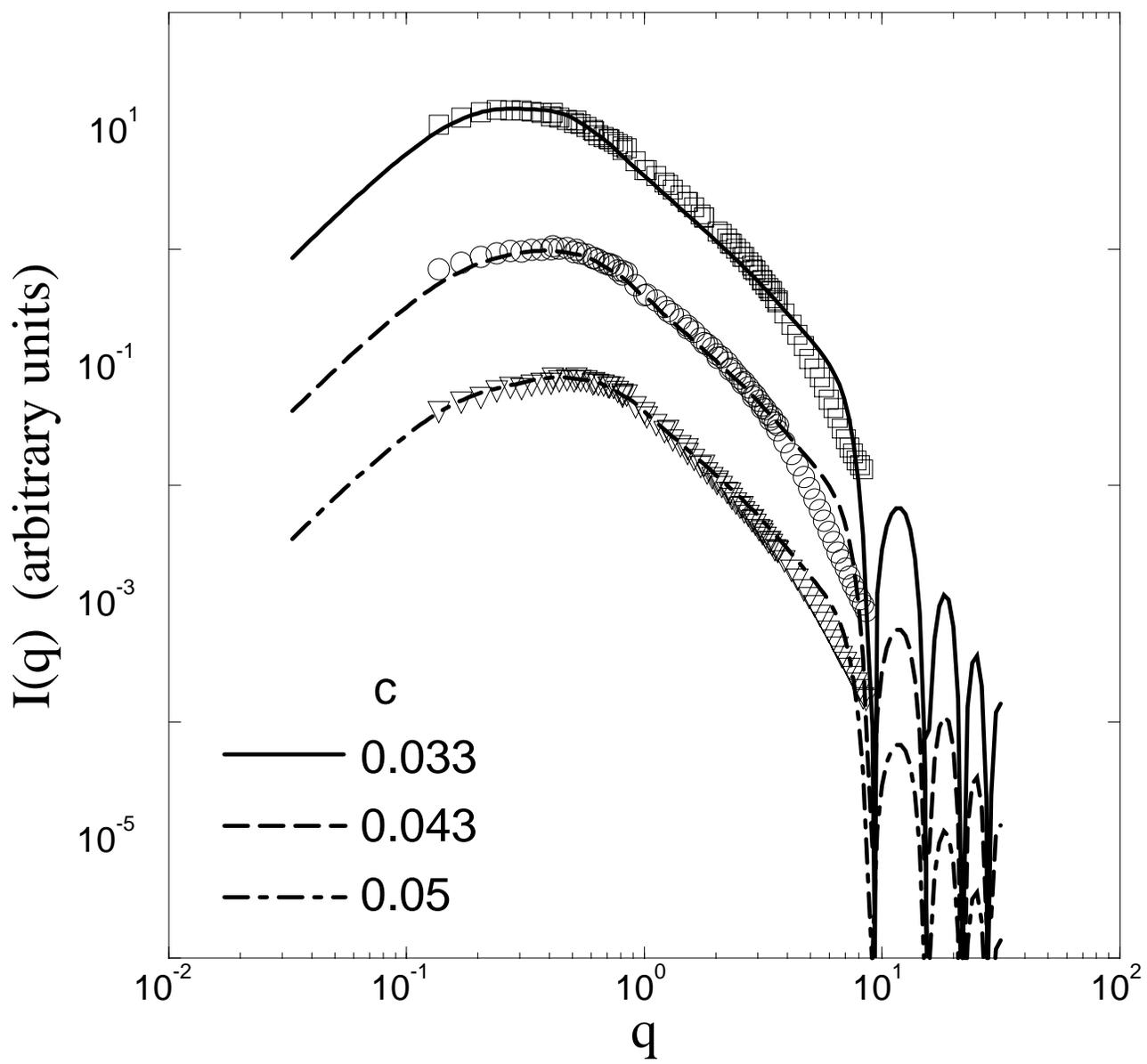
\begin{figure}
\caption{Comparison between simulations and experiments for
three base catalyzed aerogels of the same family. The concentrations
used in the simulations
$c = 0.033, 0.043, 0.05$ are calculated from the aerogel densities
($\rho = 0.073, 0.095, 0.110$g.cm$^{-3}$) using formula (11). The
two adjustable parameters,
which are a multiplicative constant for the intensity and the particles
diameter, taken to be $46\AA$, are the same for the three curves.
The curves have
been arbitrarily shifted vertically for clarity.}
\end{figure}

\begin{figure}
\caption{(a)Log-log plot of $I(q)q^4$ vs. $q$ for $\sigma_{eff}=0$
and $c=0.043$(dashed line),
and for $\sigma_{eff}=0.25$ with $N=32$ (solid line).
The open circles represent the
experimental curve for $c=0.043$.
(b) Same data as in (a) but with a more extended $q$ interval and
represented in a log-log plot of $I(q)$ vs. $q$.}
\end{figure}
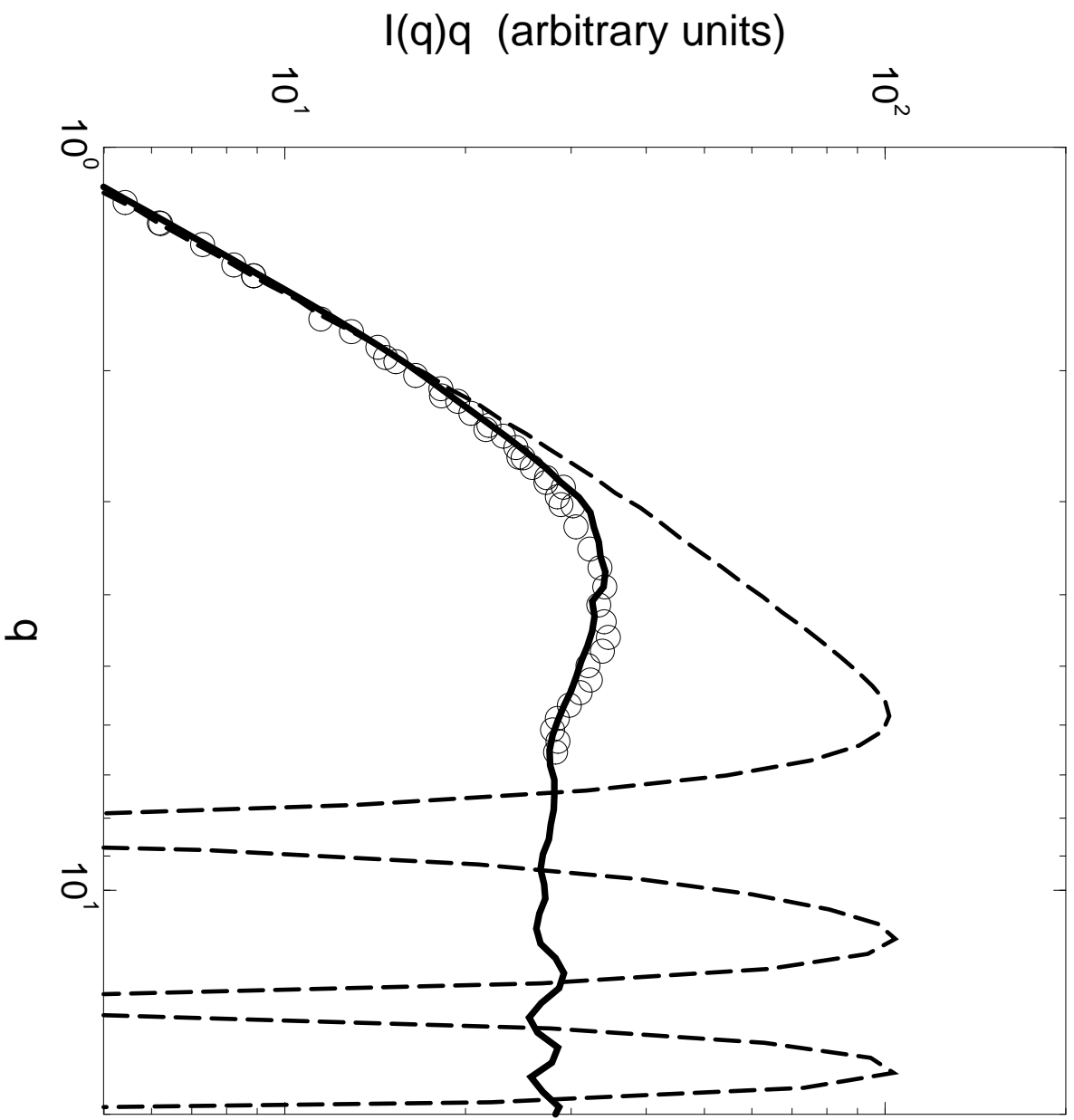
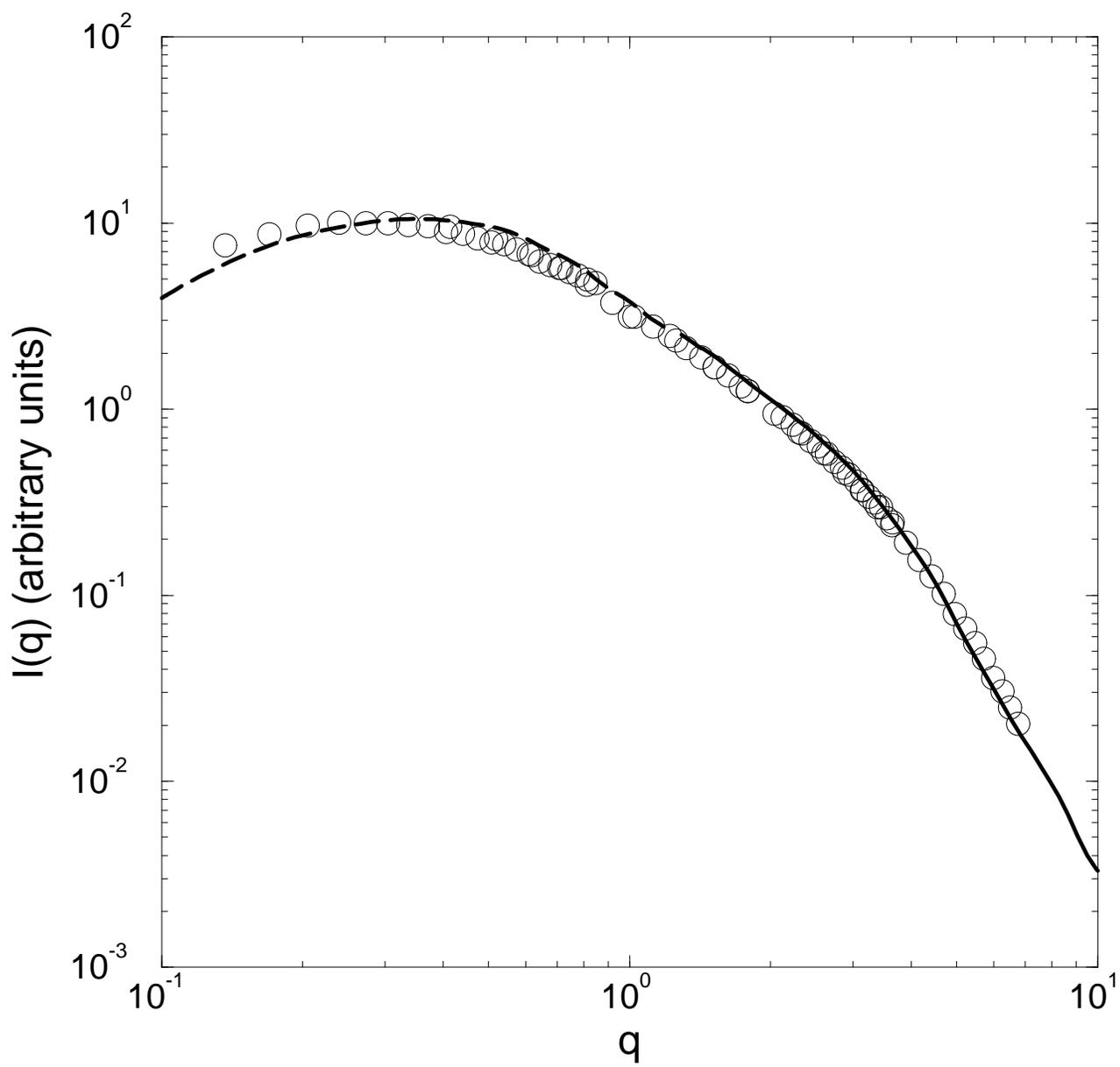
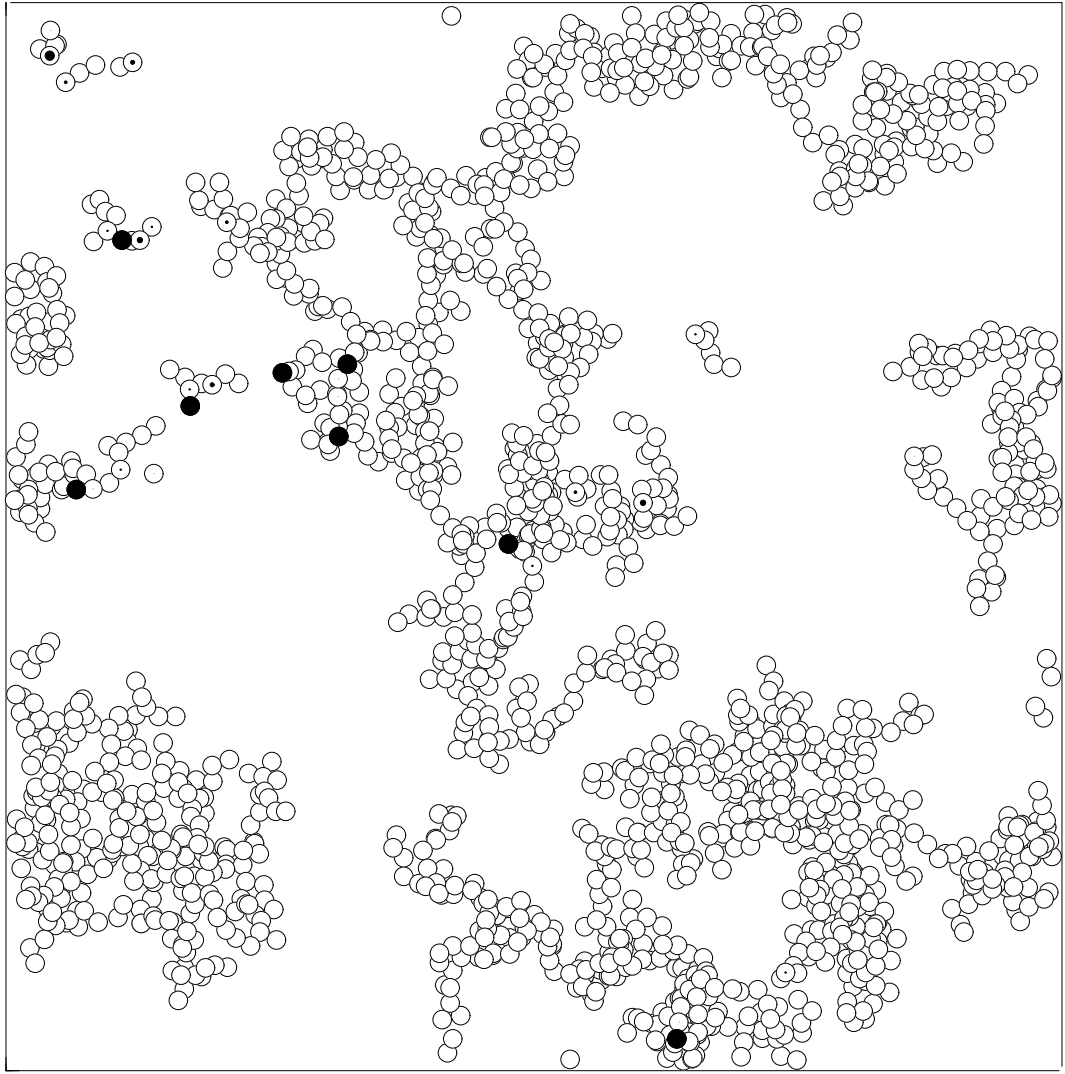
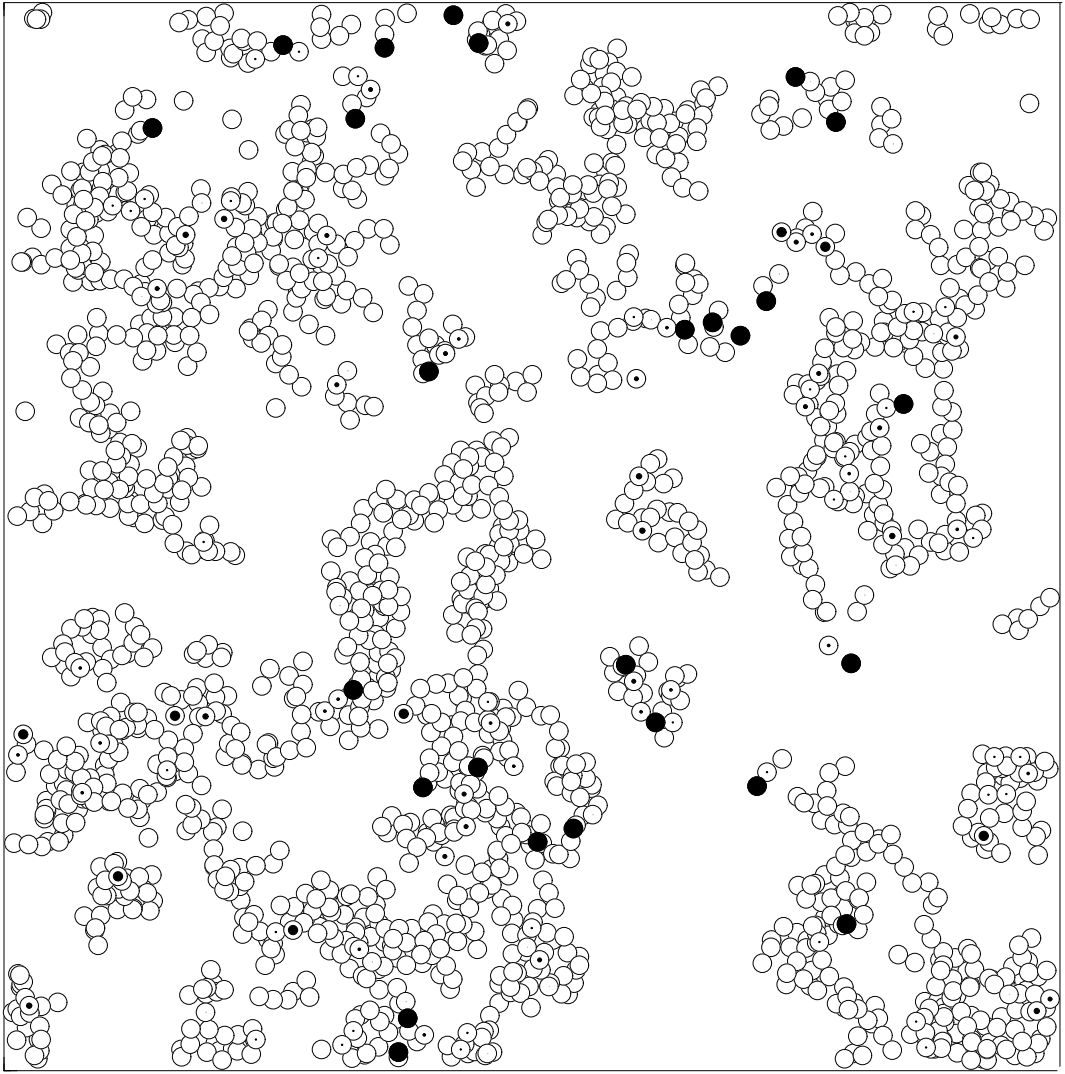
\end{document}